\documentclass[pre,aps,twocolumn,superscriptaddress,showpacs]{revtex4}
\usepackage{amssymb}
\usepackage{epsfig}
\usepackage{amsmath}
\usepackage{subfigure}
\usepackage{graphicx}
\usepackage{textcomp}
\usepackage{url}
\usepackage{float}
\usepackage{color}
\usepackage{cases}
\usepackage{color}

\begin{document}
\title{Short-time height distribution in 1d KPZ equation: starting from a parabola}

\author{Alex Kamenev}
\email{kamenev@physics.umn.edu}
\affiliation{Department of Physics, University of Minnesota, Minneapolis, MN 55455, USA}
\affiliation{William I. Fine Theoretical Physics Institute, University of Minnesota,
Minneapolis, MN 55455, USA}

\author{Baruch Meerson}
\email{meerson@mail.huji.ac.il}
\affiliation{Racah Institute of Physics, Hebrew
  University of Jerusalem, Jerusalem 91904, Israel}

\author{Pavel V. Sasorov}
\email{pavel.sasorov@gmail.com}
\affiliation{Keldysh Institute of Applied Mathematics, Moscow 125047, Russia}


\begin{abstract}

We study the probability distribution $\mathcal{P}(H,t,L)$
of the surface height $h(x=0,t)=H$ in the
Kardar-Parisi-Zhang (KPZ) equation in $1+1$ dimension when starting from a parabolic interface, $h(x,t=0)=x^2/L$.
The limits of $L\to\infty$ and $L\to 0$ have been recently solved exactly for any $t>0$.  Here we address the early-time behavior of $\mathcal{P}(H,t,L)$ for general $L$. We employ the weak-noise theory - a variant of WKB approximation -- which
yields the optimal history of the interface, conditioned on
reaching the given height $H$ at the origin at time $t$.  We find that  at small $H$
$\mathcal{P}(H,t,L)$ is Gaussian, but its tails are non-Gaussian and highly asymmetric. In the leading order and in a proper moving frame, the tails behave
as $-\ln \mathcal{P}= f_{+}|H|^{5/2}/t^{1/2}$ and $f_{-}|H|^{3/2}/t^{1/2}$. The factor $f_{+}(L,t)$ monotonically increases as a function of $L$, interpolating between time-independent values at $L=0$ and $L=\infty$ that were previously known. The factor $f_{-}$ is independent of $L$ and $t$, signalling universality of this tail for a whole class of deterministic initial conditions.

\end{abstract}

\maketitle

\section{Introduction}

The Kardar-Parisi-Zhang (KPZ) equation \cite{KPZ} describes an important universality class
of non-equilibrium interface growth \cite{HHZ,Krug,Corwin,QS,S2016}.
In $1+1$ dimension the KPZ equation,
\begin{equation}\label{KPZoriginal}
\partial_{t}h=\nu \partial^2_{x}h+(\lambda/2)\left(\partial_{x}h\right)^2+\sqrt{D}\,\xi(x,t),
\end{equation}
governs the evolution of the interface height  $h(x,t)$ driven by a Gaussian white noise  $\xi(x,t)$ with zero mean
and
$\langle\xi(x_{1},t_{1})\xi(x_{2},t_{2})\rangle = \delta(x_{1}-x_{2})\delta(t_{1}-t_{2})$. Without losing generality, we will assume that $\lambda<0$ \cite{signlambda}.

An extensive body of work was devoted to the long-time behavior of the KPZ
interface \cite{HHZ,Krug}. In $1+1$ dimension, the interface width grows at long times as $t^{1/3}$, whereas the correlation length grows as $t^{2/3}$, as confirmed in
experiments \cite{experiment1}. The exponents $1/3$ and $2/3$ are hallmarks of the KPZ universality class.
In the recent years the focus of interest in the KPZ equation in $1+1$ dimension shifted toward the complete one-point
probability distribution of height $H$ at a specified point in space and at a specified time \cite{Corwin,QS,S2016}. Several groups derived exact representations of this distribution [that we will call ${\mathcal P}(H,t,L)$]
for an arbitrary time $t>0$. This remarkable progress has been achieved for three classes of initial conditions (and some
of their combinations and variations): flat interface \cite{CLD}, sharp wedge \cite{SS,CDR,Dotsenko,ACQ,Corwin}, and stationary interface: a two-sided Brownian
interface pinned at a point \cite{IS,Borodinetal}.  In the long-time limit, and for \emph{typical} fluctuations, ${\mathcal P}(H,t)$ converges to the Gaussian orthogonal ensemble (GOE) Tracy-Widom distribution \cite{TW} for the flat interface, to the Gaussian unitary ensemble (GUE) Tracy-Widom distribution for the sharp wedge, and to the Baik-Rains distribution \cite{BR} for the stationary interface. A series of ingenious experiments fully confirmed the long-time results  \cite{experiment2}.

Recently, Le Doussal et al used the exact results for the sharp-wedge initial condition to extract asymptotics corresponding to \emph{large deviations} of $H$ at long \cite{DMS} and short \cite{DMRS} times. The long-time regime has traditionally attracted great interest \cite{HHZ,Krug,Corwin,QS,S2016}, but the short-time regime is also interesting \cite{KK2007,KK2008,KK2009,Gueudre}. Indeed, at short times one observes, for both flat and sharp-wedge initial conditions, crossover of the full one-point height statistics from the Edwards-Wilkinson universality class to the KPZ
universality class as one moves away from the body of the distribution ${\mathcal P}(H)$ to its strongly asymmetric tails \cite{KK2007,KK2009,Gueudre,MKV,DMRS}.

In each of the exactly solved cases, ${\mathcal P}(H,t)$ is given in terms of
a generating function that involves a complicated determinant form. Extracting useful asymptotics from these general results may require considerable effort. It can be advantageous to use approximations which directly probe the desired asymptotic regimes. This approach was taken in Refs. \cite{KK2007,KK2008,KK2009,MKV} which studied the short-time asymptotics of ${\mathcal P}(H,t)$ when starting the process from a flat interface. In these works the probability distribution ${\mathcal P}(H,t)$
was evaluated by using the weak-noise theory (WNT) of Eq.~(\ref{KPZoriginal}). The WNT is a variant of
WKB approximation.   It employs in a smart way the smallness of \emph{typical} noise when studying large fluctuations.
The WNT originated from the Martin-Siggia-Rose path-integral formalism in physics \cite{MSR} and the Freidlin-Wentzel large-deviation theory in mathematics \cite{FW}. The WNT is related to the optimal fluctuation method which goes back to Refs. \cite{Halperin,Langer,Lifshitz}, see also Ref. \cite{LGP}.  Similar approaches have been applied, under different names, to turbulence \cite{turb1,turb2,turb3}, stochastic reactions \cite{EK,MS2011}, diffusive lattice gases \cite{MFTreview}, and non-equilibrium surface growth \cite{Fogedby1998,Fogedby1999,Fogedby2009,KK2007,KK2008,KK2009,MV,MKV} including the KPZ equation itself. The WNT equations can be formulated as a classical Hamiltonian field theory. After having solved the WNT equations, one can evaluate the action functional, which gives, up to a sub-leading prefactor, the probability to observe a specific large deviation. The exactly soluble cases  of the complete height statistics of the KPZ equation
serve as excellent benchmarks for the WNT, which then can be applied to other initial conditions, to higher dimensions, and to other
models, where exact solutions are unavailable. Here we consider one such initial condition: a parabolic interface
\begin{equation}\label{parabola}
h(x,t=0)=\frac{x^2}{L}.
\end{equation}
The limit of $L\to\infty$ corresponds to the exactly soluble case of the flat interface. As we explain in Section 2, the limit of $L\to 0$ is intimately related to  another exactly soluble case: of the sharp wedge interface. Here we address the early-time behavior of $\mathcal{P}(H,t,L)$ for arbitrary $L$.
To this end, we determine the optimal (the most likely) history of the interface
$h(x,t)$ conditioned on reaching the height $H$ at time $T$.
We find that the tails of
$\mathcal{P}$ behave, in a proper moving frame \cite{displacement},  as $-\ln \mathcal{P}=f_+ H^{5/2}/T^{1/2}$
as $H\to \infty$
and $f_- |H|^{3/2}/T^{1/2}$ as $H\to -\infty$. The factor $f_{+}(L,T)$ increases with $L$, interpolating between previously known, time-independent values at $L=0$ and $L=\infty$. On the contrary, the factor $f_{-}$ is independent of $L$
and $T$. This indicates universality of this tail for a whole class of deterministic initial conditions, and we uncover the mechanism of, and the condition for, this universality.

Here is a plan of the remainder of this paper. In Sec. 2 we formulate the problem, identify the scaling behavior of $\mathcal{P}(H,L,T)$ and briefly discuss the connection between the problem with parabolic initial condition (\ref{parabola}) and the problem with a sharp-wedge initial condition. Our main results are presented in Sec. 3, where we employ the WNT are obtain leading-order analytical results for $-\ln {\mathcal P}(H,T,L)$ in three limiting cases:
large positive $H$, large negative $H$ and small $H$.  Section 4 contains a summary and discussion of our results.

\section{Formulation of the problem}

Without noise, the interface height is governed by the deterministic KPZ equation,
\begin{equation}\label{KPZdeterm}
\partial_{t}h=\nu \partial^2_{x}h+(\lambda/2)\left(\partial_{x}h\right)^2.
\end{equation}
Its solution with the initial condition (\ref{parabola}) is
\begin{equation}\label{determ}
h(x,t)=\frac{x^2}{L-2\lambda t}+\frac{\nu}{\lambda} \,\ln \frac{L}{L-2\lambda t},
\end{equation}
so the average profile remains parabolic at all times.  For $\lambda<0$ it is well-behaved at any $t>0$.  Let us rescale $t$ by the given time $T$ (see below), $x$ by the diffusion length $\sqrt{\nu T}$, and $h$ by the $\nu/|\lambda|$. Then Eq.~(\ref{KPZdeterm}) becomes
\begin{equation}\label{KPZdetermscaled}
\partial_{t}h=\partial^2_{x}h-(1/2)\left(\partial_{x}h\right)^2,
\end{equation}
while its solution (\ref{determ}) becomes
\begin{equation}\label{determscaled}
h_0(x,t)=\frac{x^2}{L+2 t}+\ln \left(1+\frac{2t}{L}\right),
\end{equation}
where $L$ is rescaled by $|\lambda|T$. When the rescaled $L$ is very small, the deterministic solution rapidly becomes
\begin{equation}\label{determ1}
h_0(x,t)\simeq \frac{x^2}{2 t}+\ln \left(\frac{2t}{L}\right).
\end{equation}
A very similar deterministic profile appears in the problem of sharp wedge, when $h(x,t=0)=|x|/\delta$ with
$\delta\ll 1$. Here at $t\gg \delta^2$ and $|x|\ll t/\delta$ a parabolic profile develops:
\begin{equation}\label{determwedge}
h_0(x,t)\simeq \frac{x^2}{2 t}+\ln \left(\frac{t}{\delta^2}\right).
\end{equation}
As one can see, the solutions (\ref{determ1}) and (\ref{determwedge}) are identical
up to notation. Therefore, we will not distinguish in the following between the limit of $L\to 0$ of the
parabolic initial condition and the limit of $\delta \to 0$ of the wedge initial condition.

Now we return to the stochastic equation (\ref{KPZoriginal}) and study the probability distribution  $\mathcal{P}(H,T,L)$ of observing (in a proper moving frame \cite{displacement}) a given value $h(x=0,t=T) =H$, considerably different from the prediction of the deterministic solution (\ref{determ}).
Upon the rescaling transformation introduced above, Eq.~(\ref{KPZoriginal}) becomes
\begin{equation}\label{KPZrescaled}
\partial_{t}h=\partial^2_{x}h-(1/2) \left(\partial_{x}h\right)^2+\sqrt{\epsilon} \,\xi(x,t),
\end{equation}
where
\begin{equation}\label{epsilon}
\epsilon=\frac{D\lambda^2 \sqrt{T}}{\nu^{5/2}}
\end{equation}
is a dimensionless noise magnitude. The rescaled initial condition coincides with Eq.~(\ref{parabola}), with $L$ replaced by $\tilde{L} = L/(|\lambda| T)$. As one can see,   $\mathcal{P}(H,T,L)$ depends on three dimensionless parameters: $\tilde{H}=|\lambda| H /\nu$, $\tilde{L}$ and $\epsilon$.  We will omit the tildes.

\section{Weak-noise theory}

Formally, the WNT relies on the smallness of $\epsilon$. In view of Eq.~(\ref{epsilon}), this makes the WNT especially suitable  for short times. A saddle-point evaluation of the
path integral, corresponding to Eq.~(\ref{KPZrescaled}), leads to a variational problem for the action \cite{Fogedby1998,KK2007,KK2008,KK2009,MKV}. As we show in the Appendix, the Euler-Lagrange equations can be presented as a pair of
Hamilton equations
for the optimal height history $h(x,t)$ and the canonically conjugate ``momentum" density field $\rho(x,t)$:
\begin{eqnarray}
  \partial_{t} h &=& \delta \mathcal{H}/\delta \rho = \partial_{x}^2 h -(1/2) \left(\partial_x h\right)^2+\rho ,  \label{eqh}\\
  \partial_{t}\rho &=& - \delta \mathcal{H}/\delta h = - \partial_{x}^2 \rho - \partial_x \left(\rho \partial_x h\right) ,\label{eqrho}
\end{eqnarray}
where $\mathcal{H} = \int dx \,w$ is the Hamiltonian, and  $w(x,t)= \rho\left[\partial_x^2 h-(1/2) \left(\partial_x h\right)^2+\rho/2\right]$. Note that $\rho$ undergoes rescaling  $|\lambda| T \rho/\nu \to \rho$.
The initial condition is Eq.~(\ref{parabola}) with rescaled $L$. The behavior of $h(x,t)$ at large $|x|$ is governed by
Eq.~(\ref{determ}), whereas $\rho(|x| \to \infty)=0$ so that the action is bounded, see Eq.~(\ref{action1}) below.
Finally, the condition  $h(x=0,t=1)=H$ translates into \cite{KK2007,MKV}
\begin{equation}\label{pT}
    \rho(x,1)=\Lambda \,\delta(x),
\end{equation}
where $\Lambda$ is ultimately determined by the rescaled $H$ and $L$. Once the
WNT problem is solved, we can evaluate
\begin{eqnarray}
-\ln \mathcal{P}(H,T,L)&\simeq& \frac{1}{\epsilon}\,S\left(\frac{|\lambda| H}{\nu},  \frac{L}{|\lambda|T}\right)
\nonumber \\
 &=& \frac{\nu^{5/2}}{D\lambda^2\sqrt{T}}\,\,S\left(\frac{|\lambda| H}{\nu}, \frac{L}{|\lambda|T}\right),
 \label{actiondgen}
\end{eqnarray}
(in the physical units), where the rescaled action $S$ is
\begin{equation}
S = \int_0^1 dt \int  dx\,(\rho \partial_t h -w)= \frac{1}{2}\int_0^1 dt \int  dx\,\rho^2 (x,t). \label{action1}
\end{equation}
Now we consider three asymptotic limits where we can solve the problem analytically.

\subsection{Large positive heights}

Here one can neglect the diffusion terms in Eqs.~(\ref{eqh}) and (\ref{eqrho}) and obtain hydrodynamic equations
\begin{eqnarray}
 \partial_t \rho +\partial_x (\rho V)&=& 0, \label{rhoeq}\\
  \partial_t V +V \partial_x V &=&\partial_x \rho, \label{Veq}
\end{eqnarray}
where $V(x,t) =\partial_x h (x,t)$. These equations describe a non-stationary inviscid flow of a compressible gas with density $\rho$,
velocity $V$, and \emph{negative} pressure $p(\rho)=-\rho^2/2$ \cite{MKV,MS}.  The problem should be solved
subject to the condition
\begin{equation}\label{V(x,0)}
V(x,t=0)=\frac{2x}{L}
\end{equation}
and Eq.~(\ref{pT}). Equations~(\ref{rhoeq}), (\ref{Veq}) and  (\ref{V(x,0)}) remain invariant under inviscid rescaling $x/\Lambda^{1/3} \to x$, $V/\Lambda^{1/3} \to V$, and $\rho/\Lambda^{2/3} \to \rho$. In its turn, Eq.~(\ref{pT}) becomes
\begin{equation}\label{pT1}
\rho(x,t=1)=\delta(x).
\end{equation}
Now
Eq.~(\ref{action1}) yields
\begin{equation}\label{ss1}
S=\Lambda^{5/3}\, s(L),
\end{equation}
where, in the newly rescaled variables,
\begin{equation}
s(L) = \frac{1}{2}\int_0^1 dt \int  dx\,\rho^2 (x,t). \label{action2}
\end{equation}
What is the expected scaling behavior  of $S$ entering Eq.~(\ref{actiondgen})? The rescaled height at $t=1$ is
$h(x=0,t=1) \equiv H_1(L) = H/\Lambda^{2/3}$. Therefore, $\Lambda = (H/H_1)^{3/2}$,
and Eq.~(\ref{ss1}) yields
\begin{equation}\label{svsf}
S(H,L)=\frac{s(L) H^{5/2}}{\left[H_1(L)\right]^{5/2}}
\end{equation}
leading, for any $L$, to a $H^{5/2}$ tail. What is left is to find $s(L)$ and $H_1(L)$. By virtue of the special boundary conditions (\ref{V(x,0)}) and (\ref{pT1}), the solution of Eqs.~(\ref{rhoeq}) and (\ref{Veq}) with $\rho>0$ has compact support and describes
a uniform-strain flow:
\begin{equation}
V(x,t)=a(t)\,x, \quad |x|\leq \ell(t), \label{Vin}
\end{equation}
and
\begin{numcases}
{\!\!\rho(x,t) =}
r(t) \left[1-x^2/\ell^2(t)\right], & $|x|\leq \ell(t)$, \label{rhoin}\\
0, &$|x|> \ell(t)$, \label{rhoout}
\end{numcases}
where the functions $r(t)>0$, $\ell(t)\geq 0$ and $a(t)$ are to be determined. The ``zero-pressure" region of $|x|>\ell(t)$ needs to be considered separately.

For the flat interface, $L\to \infty$, this problem was solved previously in Ref. \cite{MKV}, see also Ref. \cite{KK2009}. In that case $a(t)$ starts from zero at $t=0$ and decreases monotonically, going to $-\infty$ at $t\to 1$.  The solution describes an inflow of the gas, culminating in its collapse
into the origin at $t=1$. For a finite $L$ one has $a(t=0)=2/L>0$ [see Eq.~(\ref{V(x,0)})] implying an outflow of the gas. This outflow stops at some time $0<t_*<1$, so that $a(t_*)=0$, and then becomes an inflow, $a(t)<0$ at $t>t_*$, until $a$ reaches $-\infty$, and the gas collapses into the origin, at $t=1$.

\subsubsection{$L\to 0$}

\begin{figure}[ht]
\includegraphics[width=0.35\textwidth,clip=]{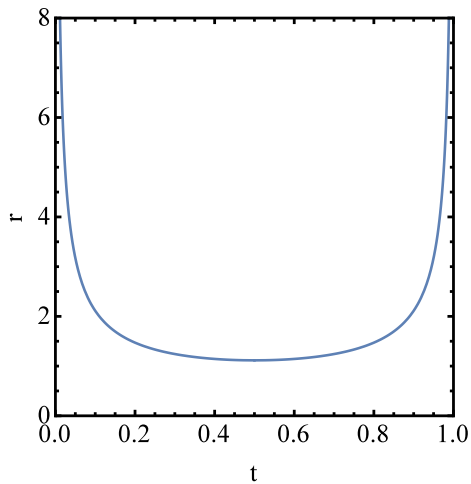}
\caption{$r=\rho(x=0,t)$ as a function of time for $H\gg 1$ and $L\to 0$ as determined by Eq.~(\ref{tplusminus}).}
\label{r(t)droplet}
\end{figure}

Let us first consider the limit of $L\to 0$ corresponding to the sharp-wedge initial condition. Here  $a(t)$ is equal to $-\infty$ at $t=0$, zero at $t=1/2$, and $+\infty$ at $t=1$. This outflow-inflow solution exhibits a remarkable symmetry in time around $t=1/2$. Here the ``gas density" $\rho$ is equal to $\delta(x)$ both at $t=0$ and at $t=1$. The  mass conservation yields  $\ell(t) r(t)=3/4$. Using it, and
plugging Eqs.~(\ref{Vin}) and~(\ref{rhoin}) into Eqs.~(\ref{rhoeq}) and (\ref{Veq}), we obtain
two coupled equations for $r(t)$ and $a(t)$: $\dot{r}=-ra$ and $\dot{a}=-a^2-(32/9) r^3$ \cite{MKV}.
Their first integral can be written as
$a=\pm (8/3) r \sqrt{r-r_*}$, where $r_*\equiv r(t=1/2)$.  This yields
\begin{equation}\label{pm}
\dot{r}=\pm (8/3) r^2 \sqrt{r-r_*},
\end{equation}
with the minus sign for $0<t<1/2$ and the plus sign for $1/2<t<1$.  An implicit solution of Eq.~(\ref{pm}),
obeying the conditions $r(t\to 0)=r(t\to 1) =\infty$ (see Fig. \ref{r(t)droplet}), is
\begin{equation}\label{tplusminus}
t=t_{\pm}(r)= \frac{1}{2}\pm \frac{3 \sqrt{r-r_*}}{8 r r_*}\pm\frac{1}{\pi}
   \arctan \left(\sqrt{\frac{r}{r_*}
   -1}\,\right),
\end{equation}
where $r_*=(3 \pi/8)^{2/3}$. The minus signs correspond to $0< t\leq 1/2$, the plus signs to $1/2\leq t< 1$.
Now we can calculate $s$:
\begin{eqnarray}
 \!\!\! s(L\to 0) &=& \!\! \frac{1}{2}\int_0^1 dt \int_{-\ell}^{\ell}  dx\,r^2(t) \left[1-x^2/\ell(t)^2\right]^2 \nonumber\\
  \!\!\!&=& \!\!\frac{2}{5} \int_0^1 dt \,r(t) = \frac{2}{5}\int_0^{1/2} r(t) dt+ \int_{1/2}^1 r(t) dt \nonumber\\
  \!\!\!&=&\!\!\frac{2}{5} \left(\int_{\infty}^{r_*} dr \, r\frac{dt_{-}}{dr}+\int_{r_*}^{\infty} dr \, r\frac{dt_{+}}{dr}\right)
  \nonumber \\
  \!\!\!&=& \!\frac{(3 \pi)^{2/3}}{5}.
\label{s1result}
\end{eqnarray}
To determine $H_1$, we can use Eq.~(\ref{eqh}) at $x=0$:
\begin{equation}\label{KPZ0}
\partial_{t}h(0,t)=\partial_{x}^2 h(0,t) -\frac{1}{2} \left[\partial_x h\right(0,t)]^2+\rho(0,t).
\end{equation}
As $\partial_x h(0,t)=0$ (except at $t=0$ and $t=1$), and the diffusion is negligible, we obtain
\begin{equation}\label{KPZ00}
\partial_{t}h(0,t) \simeq \rho(0,t) = r(t),
\end{equation}
so
\begin{equation}\label{H1L=0}
H_1= \int_0^1 r(t) dt = \frac{(3\pi)^{2/3}}{2}.
\end{equation}
Now we plug $s$ and $H_1$ into Eq.~(\ref{svsf}) and obtain the $H\gg 1$ tail we are after.
In the physical units,
\begin{equation}\label{invisciddroplet}
-\ln \mathcal{P}(H,T, L\to 0) \simeq \frac{4\sqrt{2 |\lambda|}}{15 \pi D}\, \frac{H^{5/2}}{T^{1/2}}.
\end{equation}
Equation~(\ref{invisciddroplet}) coincides with
the asymptotic (5) of Ref. \cite{DMRS,units}, extracted from the exact solution \cite{SS,CDR,Dotsenko,ACQ,Corwin} at short times. This leading-order asymptotic is controlled by the nonlinearity and independent of $\nu$.  It is twice as small as the corresponding result  \cite{KK2009,MKV} for $L\to \infty$.

\begin{figure}[ht]
\includegraphics[width=0.4\textwidth,clip=]{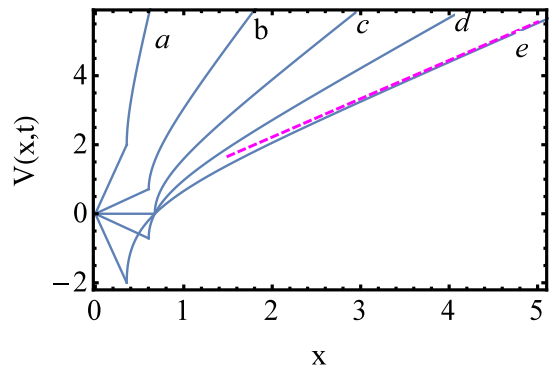}
\caption{The rescaled interface slope $V(x,t)=\partial_x h(x,t)$, as described by the inviscid solution for $H\gg 1$, is shown as a function of $x>0$ at times $t=0.1$ (a), $0.3$ (b), $0.5$ (c), $0.7$ (d) and $0.9$ (e) for $L\to 0$. Both the ``pressure-driven" solution and the Hopf solution are shown. A stagnation point $V=0$ develops at $r=r_*$ at $t\geq 1/2$. The interface height $h(x,t)$ has a local minimum at this point at all times $t\geq 1/2$. The dashed line is the large-$x$ asymptotic $V=x/t$ at $t=0.9$.}
\label{VforL=0}
\end{figure}

In the zero-pressure region $|x|>\ell(t)$ the governing equation,
\begin{equation}\label{P010}
\partial_t V + V\partial_x V=0,
\end{equation}
describes the Hopf flow. We will only consider $x>\ell(t)$; the solution for $x<-\ell(t)$ can be obtained from the symmetry
$V(-x,t)=-V(x,t)$. The general solution of Eq.~(\ref{P010}) can be written as \cite{LL,Whitham}
\begin{equation}\label{P020}
x-Vt=F(V)\,,
\end{equation}
where the arbitrary function $F(V)$ should be found from
matching with the pressure-driven solution at $x=\ell(t)$. The matching yields the equation
\begin{equation}\label{P050}
x-Vt=\frac{3}{4r_*}-\frac{V}{2}+\frac{V}{\pi}\,\arctan \frac{V}{2\sqrt{r_*}}
\end{equation}
which determines $V(x,t)$ in an implicit form. Figure \ref{VforL=0} shows $V$ as a function of $x>0$ at different times. Both the pressure-driven solution (\ref{Vin}) and the Hopf solution (\ref{P050}) are shown. Importantly, the Hopf solution
complies with the large-$x$ asymptotic $V(x,t) \simeq x/t$, described by the inviscid limit of the deterministic solution (\ref{determ}) at $L\to 0$. Notice the presence of the stagnation point at $r=r_*$ at $t\geq 1/2$.  We also calculated $h(x,t)$ in an implicit form, but we do not present these cumbersome formulas here.

\subsubsection{$L>0$}

\begin{figure}[ht]
\includegraphics[width=0.45\textwidth,clip=]{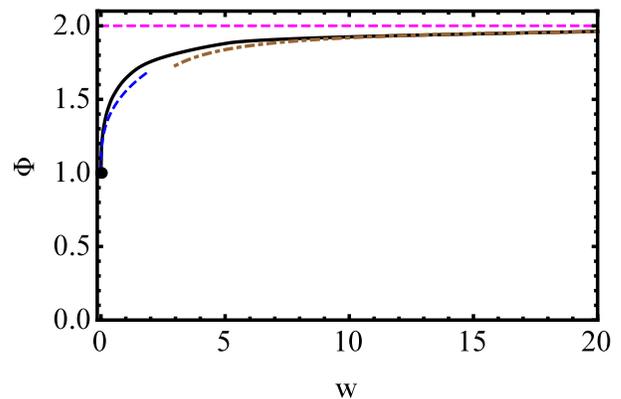}
\caption{$\Phi(w)$ from Eq.~(\ref{inviscidactgeneral}) and its asymptotics, Eqs.~(\ref{fsmall}) and (\ref{flarge}). The filled circle shows $\Phi(0)=1$,  the horizontal asymptotic shows $\Phi(\infty)=2$.
}
\label{Phi}
\end{figure}

In this case $a(t=0)=2/L$, $a(t=1)=-\infty$, and $a(t=t_1)=0$ where $0<t_1<1$ is a priori unknown. Let us denote $r(t_1)=r_1$. The first integral of the equations for $\dot{a}$ and $\dot{r}$ can be written as $a=\pm (8/3) r \sqrt{r-r_1}$ leading to
\begin{equation}\label{pmgen}
\dot{r}=\pm (8/3) r^2 \sqrt{r-r_1}.
\end{equation}
An implicit solution of Eq.~(\ref{pmgen} is
\begin{equation}\label{tplusminusgen}
t=t_{\pm}(r)= t_1\pm\frac{3}{8} \left(\frac{\sqrt{r-r_1}}{r r_1}+\frac{\arctan \sqrt{\frac{r}{r_1}-1}}{r_1^{3/2}}\right),
\end{equation}
where $r_0\equiv r(t=0)$ is a priori unknown. In Eqs.~(\ref{pmgen}) and (\ref{tplusminusgen}) the minus signs correspond to $0<t<t_1$, the plus signs to $t_1<t<1$.
Let us evaluate the rescaled action:
\begin{eqnarray}
 \!\!\! s &=&\!\!\frac{2}{5} \int_0^1 dt \,r(t) = \frac{2}{5}\int_0^{t_1} r(t) dt+ \int_{t_1}^1 r(t) dt \nonumber\\
  \!\!\!&=&\!\!\frac{2}{5} \left(\int_{r_0}^{r_1} dr \, r\frac{dt_{-}}{dr}+\int_{r_1}^{\infty} dr \, r\frac{dt_{+}}{dr}\right) \nonumber \\
  \!\!\!&=&\!\!\frac{3 \left(\pi+2 \arccos \sqrt{\frac{r_1}{r_0}}\right)}{20 \sqrt{r_1}}.
\label{s1resultgen}
\end{eqnarray}
Also,
\begin{equation}\label{H1gen}
H_1 \simeq \int_0^1 r(t)\,dt =\frac{3 \left(\pi+2 \arccos \sqrt{\frac{r_1}{r_0}}\right)}{8 \sqrt{r_1}}.
\end{equation}
The three unknown constants $r_0$, $r_1$ and $t_1$ can be expressed via $L$, the only parameter of the rescaled problem, with the help of three algebraic relations:
\begin{eqnarray}
  &&\frac{8}{3} r_0 \sqrt{r_0-r_1}= \frac{2}{L}, \label{relation1}\\
  &&t_1= \frac{3}{8} \left(\frac{\sqrt{r_0-r_1}}{r_0 r_1}+\frac{\arctan \sqrt{\frac{r_0}{r_1}-1}}{r_1^{3/2}}\right), \label{relation2}\\
 && t_1+\frac{3}{8}\frac{\pi}{2 r_1^{3/2}} =1. \label{relation3}
\end{eqnarray}
The solution is unique and can be obtained in a parametric form. Let us introduce the parameter $y=r_0/r_1$ that decreases monotonically from $\infty$ to $1$ as $L$ increases from $0$ to $\infty$. We can express $L, r_0, r_1$ and $t_1$ via $y$ as follows:
\begin{eqnarray}
\!\!\!\!\!\!\!\!L&=&\!\!\frac{4}{\sqrt{y-1} \left[\pi  y+2 \sqrt{y-1}+2 y \arctan\left(\sqrt{y-1}\right)\right]},
\label{P270}\\
\!\!\!\!\!\!\!\!r_0&=&\!\!\frac{3^{2/3}}{4}y\left(\frac{\pi}{2} +\frac{\sqrt{y-1}}{y}+\arctan\sqrt{y-1}\right)^{2/3},
\label{P260}\\
\!\!\!\!\!\!\!\!r_1&=&\!\! \frac{3^{2/3}}{4}\left(\frac{\pi}{2}+\frac{\sqrt{y-1}}{y}+\arctan\sqrt{y-1}\right)^{2/3},
\label{P240}\\
\!\!\!\!\!\!\!\!t_1&=&\!\!2\frac{\frac{\sqrt{y-1}}{y}+\arctan\sqrt{y-1}}{\pi +2\left(\frac{\sqrt{y-1}}{y}+\arctan\sqrt{y-1}\right)}.
\label{P250}
\end{eqnarray}
Using these relations in conjunction with Eqs. (\ref{s1resultgen}) and (\ref{H1gen}), and introducing $\Phi(L)=s_1/H_1^{5/2}$, we finally obtain, in
physical units,
\begin{equation}\label{inviscidactgeneral}
-\ln \mathcal{P}(H,T,L)\simeq \frac{4\sqrt{2}\,|\lambda|^{1/2} H^{5/2}}{15 \pi D T^{1/2}} \Phi\left(\frac{L}{|\lambda|T}\right).
\end{equation}
Correspondingly, the factor $f_+(L,T)$, mentioned in the Abstract and in the Introduction, is the following:
$$
f_+=\frac{4\sqrt{2}\,|\lambda|^{1/2}}{15 \pi D} \Phi\left(\frac{L}{|\lambda|T}\right).
$$
A plot of the function $\Phi=\Phi(w)$ is shown in Fig. \ref{Phi}. Its small- and large-$w$ asymptotics are
\begin{numcases}
{\!\!\Phi(w) \simeq} 1+\frac{3 w^{1/3}}{2^{4/3} \pi^{2/3}}, & $w\ll 1$, \label{fsmall}\\
2\left(1-\frac{4}{\pi^2 w}\right), &$w\gg 1$, \label{flarge}
\end{numcases}
see Fig. \ref{Phi}. At $L\to 0$ we obtain $\Phi=1$  (the solid point on Fig. \ref{Phi}), in agreement with Eq.~(\ref{invisciddroplet}) and Ref. \cite{DMRS}.  At $L\to \infty$ one has $\Phi = 2$ (the horizontal dashed line) in agreement with Refs. \cite{KK2009,MKV}. Notice the non-analytic $w^{1/3}$ behavior of $f(w)$ at $w\to 0$.

The Hopf flow regions $|x|>\ell(t)$ for $L>0$ can be analyzed similarly to the case of $L\to 0$.  The Hopf-flow solution for $V(x,t)$ matches continuously with the pressure-driven solution at $|x|=\ell(t)$, complies with the deterministic behavior $V(x,t)=2x/(L+2t)$ at $|x|\to \infty$, and exhibits, at $t\geq t_1$, two stagnation points $V=0$ at $x=\pm r_1$ where $h(x,t)$ has a local minimum. We do not show these cumbersome formulas here.

\subsection{Large negative heights}
At very large negative $H$, or $\Lambda$, the solution, at any $L$,  has the following character.  $\rho$ is localized in a narrow boundary layer around $x=0$ and is almost independent of time except very close to $t=0$ and $t=1$. $V$  in the boundary layer  is also almost independent of time. There is also exterior, or bulk, region where $\rho\simeq 0$, whereas $V(x,t)$ obeys the deterministic KPZ equation (\ref{KPZdeterm}).

\subsubsection{The boundary layer}

The stationary boundary-layer solution was previously found in the problem of flat interface \cite{KK2007,MKV}, see
also Ref. \cite{Fogedby1998}:
\begin{eqnarray}
\rho_{\text{bl}}(x) &=& -2 c\,\text{sech}^2 \left( \sqrt{c/2} \,x \right),  \label{pin} \\
V_{\text{bl}}(x) &=& \sqrt{2 c}\,\tanh \left(\sqrt{c/2} \,x \right), \label{Vin1}
\end{eqnarray}
where $c=\Lambda^2/32$. The action in terms of $c$ or $\Lambda$ is obtained immediately:
\begin{equation}\label{ssoliton}
S =\frac{1}{2} \int_{-\infty}^{\infty} dx\,\rho_{\text{bl}}^2(x) = \frac{8\sqrt{2}\, c^{3/2}}{3}= -\frac{\Lambda^3}{48},
\end{equation}
recall that $\Lambda<0$. To express $c$ through $H$, we need to rewrite the boundary-layer solution in terms of $h(x,t)$ \cite{MKV},
\begin{equation}\label{hbl}
h_{\text{bl}}(x,t) = 2 \ln \cosh \left( \sqrt{c/2} \,x \right) -c t,
\end{equation}
and obtain $h_{\text{bl}}(0,1) = -c =H$ which yields $c=-H$ and $\Lambda = - 2^{5/2} |H|^{1/2}$. Using this result
in Eqs.~(\ref{actiondgen}) and~(\ref{ssoliton}), we obtain
in the physical units
\begin{equation}\label{negativetail}
-\ln \mathcal{P} (H,T,L)\simeq \frac{8\sqrt{2}\,\nu |H|^{3/2}}{3 D |\lambda|^{1/2} T^{1/2}}.
\end{equation}
As one can see, the factor
$$
f_- = \frac{8\sqrt{2}\,\nu}{3 D |\lambda|^{1/2}}
$$
is independent of $L$ and $T$. It is not surprising, therefore, that the same expression (\ref{negativetail}) for the negative tail was previously obtained for  $L\to \infty$ \cite{KK2007,MKV} and $L\to 0$ \cite{DMRS}. Interestingly, Eq.~(\ref{negativetail}) also coincides with the corresponding asymptotic of the GOE and GUE Tracy-Widom distributions,
which describe the negative tail of ${\mathcal P}(H,T)$ at \emph{long} times, both for  $L\to \infty$ \cite{CLD,KK2007,MKV} and for $L\to 0$ \cite{SS,CDR,Dotsenko,ACQ,Corwin}.

\subsubsection{The bulk region}

Now we will show that the boundary-layer solution (\ref{Vin1}) can be properly matched with a deterministic bulk solution.
Not being interested in the structure of an additional narrow transition layer that emerges in the bulk solution (see below), we can neglect the diffusion term and, instead of Eq.~(\ref{KPZdeterm}),  deal with the inviscid equation
\begin{equation}\label{hHopf}
\partial_t h +\frac{1}2{} (\partial_x h)^2=0,
\end{equation}
or the Hopf equation (\ref{P010}),  where we allow for $V$-shocks. We will only consider $x>0$: the solution for $x<0$ can obtained by a mirror reflection of $h(x,t)$ with respect to the origin. The outer asymptotic of the boundary-layer solution (\ref{hbl}) for $h(x,t)$ is
\begin{equation}\label{h1}
h_1(x,t) = \sqrt{2c} \,x - ct.
\end{equation}
Correspondingly, $V_1(x,t)=\partial_x h_1(x,t)=\sqrt{2c}=|\Lambda|/4=\text{const}$. Note that these asymptotics is independent of the diffusivity. To satisfy the boundary conditions at $x \to \infty$, $h_1(x,t)$ must be continuously matched with the inviscid limit of the deterministic solution (\ref{determscaled}), which holds at large distances,
\begin{equation}\label{h2}
h_2(x,t) \simeq \frac{x^2}{L+2t},
\end{equation}
and for which
\begin{equation}\label{V2}
V_2(x,t)\simeq \frac{2x}{L+2t}.
\end{equation}
At $L>0$ the equality $h_1(x,t)=h_2(x,t)$ is satisfied in two locations, $X_-(t)$ and $X_+(t)$, where
\begin{equation}\label{Xpm}
X_{\pm}(t) = \sqrt{(c/2)\left(L+t\right)}\,\left(\sqrt{L+t}\pm\sqrt{L}\right) .
\end{equation}
While $h(x)$ is continuous in the matching points, $V(x)$ is generally not, so a shock appears.
$X_+(t)$ is inadmissible as a shock position, as it violates the condition $V_1[X(t),t] \geq  V_2[X(t),t]$ \cite{Whitham}.
$X_-(t)$ does satisfy this condition, and so $V(x,t)$ exhibits a shock at this location.
The shock speed is equal to
\begin{equation}\label{shockspeed}
V_{\text{shock}}(t) = \frac{dX_-}{dt} =\sqrt{2 c}-\sqrt{\frac{c
   L}{2(L+2 t)}}.
\end{equation}
What happens in the limits of $L\to \infty$ and $L\to 0$? At $L\to \infty$ the deterministic solution at large distances is trivial: $h_2(x,t)=0$. Here the $V$-shock is
located at $X(t)=\sqrt{c/2}\,t$ and moves with a constant speed \cite{MKV}. In the limit of $L\to 0$ the two locations $X_-(t)$ and $X_+(t)$ merge.
In this special case  $V(x)$ is continuous everywhere, and there is no shock. There is only a discontinuity in the derivative $\partial_x V$ at the moving point $X(t)= \sqrt{2c}\,t$. All the discontinuities, discussed here, are smoothed, and narrow transition layers appear, if one accounts for the diffusion.

\subsection{The variance}

\begin{figure}[ht]
\includegraphics[width=0.45\textwidth,clip=]{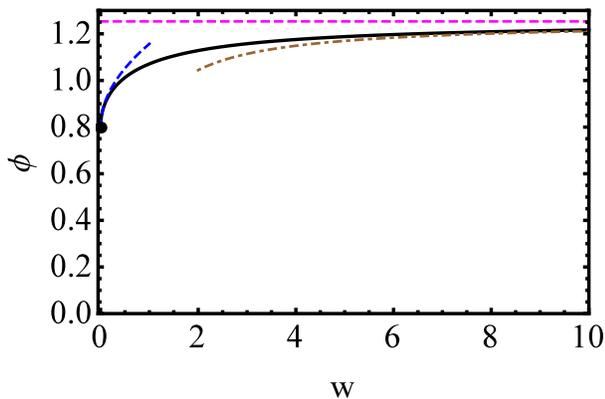}
\caption{$\phi(w)$ from Eqs.~(\ref{actiongauss}) and (\ref{phi}) and its asymptotics, Eqs.~(\ref{phismall}) and (\ref{philarge}). The filled circle shows $\phi(0)=\sqrt{2/\pi}$,  the horizontal asymptotic shows $\phi(\infty)=\sqrt{\pi/2}$.}
\label{phifig}
\end{figure}

When $\epsilon \ll 1$,  low cumulants of ${\mathcal P}$ can be calculated via a regular
perturbation theory in $H$, or in $\Lambda$, in the WNT framework \cite{MKV,KrMe}. We set
\begin{eqnarray}
 \!\!\! h(x,t)&=& h_0(x,t)+\Lambda h_1(x,t)+\Lambda^2 h_2(x,t) +\dots ,\label{hexpansion}\\
 \!\!\! \rho(x,t) &=& \Lambda \rho_1(x,t)+\Lambda^2 \rho_2(x,t) +\dots . \label{pexpansion}
\end{eqnarray}
where $h_0(x,t)$ is given by Eq.~(\ref{determscaled}).
Correspondingly, $S(\Lambda)=\Lambda^2 S_1 +\Lambda^3 S_2 + \dots$. Here we limit
ourselves to the first order of this perturbation series which gives the distribution variance.
In the first order Eqs.~(\ref{eqh}) and (\ref{eqrho})  yield
\begin{eqnarray}
  && \partial_t h_1 +\partial_x h_0 \,\partial_x h_1 - \partial_x^2 h_1 = \rho_1, \label{1}\\
  && \partial_t \rho_1 +\partial_x (\partial_x h_0 \,\rho_1)+\partial_x^2 \rho_1= 0, \label{2}
\end{eqnarray}
or
\begin{eqnarray}
   &&  \partial_t h_1 +\frac{2 x}{L+2t} \,\partial_x h_1 - \partial_x^2 h_1=\rho_1, \label{1a}\\
   &&  \partial_t \rho_1 +\partial_x \left(\frac{2 x}{L+2t} \,\rho_1\right)+\partial_x^2 \rho_1=0. \label{2a}
\end{eqnarray}
In contrast to the flat case \cite{MKV,Gueudre}, the KPZ nonlinearity kicks in already in the first  order
of the perturbation theory, so the variance of ${\mathcal P}(H,T,L)$ is different from that for
the Edwards-Wilkinson equation. To solve Eqs.~(\ref{1a}) and (\ref{2a}), we introduce new variables
$$
z=\frac{x}{L+2t},\quad u(z,t) = (L+2t) \rho_1.
$$
Equation~(\ref{2a}) becomes
\begin{equation}\label{ueq}
\partial_t u+\frac{\partial_z^2 u}{(L+2t)^2}=0.
\end{equation}
Now we introduce new time,
$$
\tau=\frac{t}{L(L+2t)},
$$
so that $t=\tau L^2/(1-2 \tau L)$. The new time grows monotonically on the interval $0\leq \tau \leq \tau_1$, where
$$
\tau_1 = \frac{1}{L(L+2)}
$$
corresponds to $t=1$.
Equation~(\ref{ueq}) becomes the antidiffusion equation $\partial_{\tau} u +\partial_{z}^2 u=0$.
The boundary condition $\rho_1(x,1)=\delta(x)$ translates into $u(z,\tau_1)=\delta(z)$,
and the solution is
\begin{equation}\label{uresult}
u(z,0\leq \tau\leq \tau_1)= \frac{1}{\sqrt{4 \pi(\tau_1-\tau)}}\,e^{-\frac{z^2}{4(\tau_1-\tau)}},
\end{equation}
or
\begin{equation}\label{p1result}
\rho_1(x,t)=\frac{e^{-\frac{(L+2) x^2}{4 (1-t)
   (L+2 t)}}}{\sqrt{\frac{4\pi\,(1-t) (L+2
   t)}{L+2}}}
\end{equation}
As a result,
\begin{eqnarray}
  S_1(L) &=&  \frac{1}{2}\int_0^1 dt \int_{-\infty}^{\infty} dx\,\rho_1^2(x,t) \nonumber \\
 &=& \frac{\sqrt{L+2} \,\arccos\left(\frac{L-2}{L+2}\right)}{8 \sqrt{\pi}}, \label{S1}
\end{eqnarray}
To express $\Lambda$ via $H$ we
need to solve Eq.~(\ref{1a}) for $h_1(x,t)$ with the initial condition $h_1(x,0)=0$ and
the source term given by Eq.~(\ref{p1result}). It suffices to calculate $h_1(x=0,t=1)$.
In the new variables Eq.~(\ref{1a}) becomes
\begin{equation}\label{1b}
\partial_{\tau} h = \partial_z^2 h+\frac{L u(z,\tau)}{1-2\tau L},
\end{equation}
with $u(z,\tau)$ from Eq.~(\ref{uresult}). The solution can be obtained
with the help of the Green's function of the diffusion equation. As a result,
\begin{eqnarray}
 && h_1(x=0, t=1)=h_1(z=0, \tau=\tau_1) \nonumber \\
  && =\frac{1}{4\pi}\int_0^{\tau_1} \frac{L d\tau}{(1-2L\tau)(\tau_1-\tau)}\,\int_{-\infty}^{\infty} dz\,e^{-\frac{z^2}{2(\tau_1-\tau)}}\nonumber \\
  && =\frac{\sqrt{L+2}}{2\sqrt{\pi}}\,\arccos \sqrt{\frac{L}{L+2}}.
\end{eqnarray}
Now we can express $\Lambda$ through $H$ using the relation $\Lambda \,h_1(x=0,t=1) = H$. Finally, we obtain in the physical units
\begin{equation}\label{actiongauss}
-\ln {\mathcal P}(H,T,L)\simeq \frac{\nu^{1/2}H^2}{D\sqrt{T}}\,\phi\left(\frac{L}{|\lambda|T}\right),
\end{equation}
where
\begin{equation}\label{phi}
\phi(w)=\frac{\sqrt{\pi}\,\arccos\left(\frac{w-2}{w+2}\right)}{2\sqrt{w+2}\,\left(\arccos\sqrt{\frac{w}{w+2}}\right)^2}.
\end{equation}
The asymptotics of $\phi(w)$ are the following:
\begin{numcases}
{\!\!\phi(w) \simeq}\sqrt{\frac{2}{\pi}}\left(1+\frac{\sqrt{2w}}{\pi}\right) , & $w\ll 1$, \label{phismall}\\
\sqrt{\frac{\pi}{2}}\left(1-\frac{1}{3w}\right), &$w\gg 1$, \label{philarge}
\end{numcases}
see Fig. \ref{phifig}.  At $L\to 0$ we obtain $\phi=\sqrt{2/\pi}$ in agreement with Eq.~(6) of Ref. \cite{DMRS,units}.
At $L\to \infty$ $\phi=\sqrt{\pi/2}$ in agreement with Refs. \cite{Gueudre,MKV}.  Notice the non-analytic $w^{1/2}$ behavior of $\Phi(w)$ at $w\to 0$.

\section{Summary and Discussion}

Let us briefly summarize our results for the probability distribution $\mathcal{P}(H,t,L)$
of the surface height $h(x=0,t)=H$ in the
KPZ equation in $1+1$ dimension when starting from a parabolic interface $h(x,t=0)=x^2/L$.

At early times, $\epsilon \ll 1$, the central part of the distribution is described by Eqs.~(\ref{actiongauss}) and (\ref{phi}). Although it is a Gaussian, it does not belong to the Edwards-Wilkinson universality class. Indeed,
the distribution variance explicitly depends on the nonlinearity coefficient $\lambda$ and does not exhibit the customary $t^{1/4}$ scaling, see Eq.~(\ref{actiongauss}).

The tails of $\mathcal{P}(H,t,L)$ are described by Eqs.~(\ref{inviscidactgeneral}) and (\ref{negativetail}): they are non-Gaussian and strongly asymmetric. The asymmetry is manifested by
very different optimal histories of the process conditioned on observing a large positive or
negative value of $H$ at time $t$.

As we observed, the positive tail (\ref{inviscidactgeneral}) of ${\mathcal P}(H,t,L)$ depends on $L$ monotonically, see Fig. \ref{Phi}, interpolating between time-independent values at $L=0$ and $L=\infty$ that were previously known. On the contrary, the negative tail (\ref{negativetail}) of ${\mathcal P}(H,t,L)$ is independent of $L$, because it comes from the universal boundary-layer solution (\ref{pin}) and (\ref{Vin1}). We argue that exactly the same negative tail (\ref{negativetail}) should be observed for a whole class of deterministic initial conditions such that the boundary-layer solution (\ref{Vin1}) for $V(x,t)=\partial_x h(x,t)$ can be matched with a (deterministic) bulk solution for $V(x,t)$
that satisfies correct boundary conditions at $|x|\to \infty$. An important role in this matching is played by $V$-shocks (in the inviscid approximation) that, in general, develop inside the bulk region.

Are any of our  early-time predictions, based on the WNT, expected to hold at long times? (See Ref. \cite{MKV} for a similar discussion for the flat initial condition.) At $\epsilon \gg 1$, the WNT breaks down in the body of the height distribution, where the Gaussian distribution (\ref{actiongauss}) and (\ref{phi}) must give way to a different distribution which reduces to the GUE Tracy-Widom statistics at $L\to 0$, and to the GOE Tracy-Widom statistics at $L\to \infty$. However, sufficiently far in the tails the action $S$ is very large. Therefore, one can  expect
the WNT tails (\ref{inviscidactgeneral})
and (\ref{negativetail}) to hold there. Indeed, the universal $3/2$ tail agrees with the corresponding Tracy-Widom tail at $L\to 0$ and $L\to \infty$. The  $5/2$ tail is incompatible with the Tracy-Widom statistics. We argue that it holds (see also Ref. \cite{MKV}) when
it predicts a higher probability than the corresponding tail, $-\ln \mathcal{P}\sim  \nu^2 H^3/(|\lambda|D^2 t)$, of the Tracy-Widom distribution. At fixed $t$, and sufficiently far in the tail, $H\gg D^2 |\lambda|^3 t/\nu^4$, this condition is satisfied. It would be very interesting to test this prediction by extracting the $H\gg D^2 |\lambda|^3 t/\nu^4$ asymptotics of  ${\mathcal P}$ in the exactly soluble cases of $L\to 0$ and $L\to \infty$.

\section*{ACKNOWLEDGMENTS}

A.K. was supported by NSF grant DMR1306734. B.M. acknowledges financial support from the Israel Science Foundation (grant No. 807/16) and the United States-Israel Binational Science
Foundation (BSF) (grant No.\ 2012145)
and hospitality of the William I. Fine Theoretical Physics Institute of the University of Minnesota.

\section*{Appendix: Derivation of the WNT Equations}

\renewcommand{\theequation}{A\arabic{equation}}
\setcounter{equation}{0}

For completeness, here we present a brief derivation of the WNT equations and boundary conditions. Using Eq.~(1), we express the noise term as
\begin{equation}\label{actn0}
\sqrt{D}\,\xi(x,t)=\partial_{t} h-\nu \partial_{x}^2 h-\frac{\lambda}{2} \left(\partial_{x} h\right)^2.
\end{equation}
The corresponding Gaussian action is, therefore,
\begin{equation}\label{actn}
S=\frac{1}{2}\int_{0}^{T}dt\int_{-\infty}^{\infty}dx \left[\partial_{t} h-\nu \partial_{x}^2 h-\frac{\lambda}{2} \left(\partial_{x} h\right)^2\right]^2.
\end{equation}
In the weak-noise limit the main contribution to the integral comes from the ``optimal path" $h(x,t)$ that minimizes $S$. The variation of $S$
\begin{eqnarray}
  \delta S&=& \int_{0}^{T}dt\int_{-L/2}^{L/2}dx\left[\partial_{t} h-\nu \partial_{x}^2 h-\frac{\lambda}{2} \left(\partial_{x} h\right)^2\right] \nonumber\\
 &\times& \left(\partial_t \delta h -\nu \partial_x^2 \delta h -\lambda \partial_x h \,\partial_x \delta h\right).
 \label{variation}
\end{eqnarray}
By analogy with classical mechanics, one can introduce the ``momentum density" field $\rho(x,t)=\delta L/\delta v$, where $v\equiv \partial_t h$, and
$$
L\{h\}=\frac{1}{2}\int_{-\infty}^{\infty}dx \left[\partial_{t} h-\nu \partial_{x}^2 h-\frac{\lambda}{2} \left(\partial_{x} h\right)^2\right]^2
$$
is the Lagrangian.  In this way we obtain
\begin{equation}\label{heqA}
\partial_{t}h=\nu \partial_{x}^2 h +\frac{\lambda}{2} \left(\partial_x h\right)^2+\rho,
\end{equation}
the first of the two Hamilton equations. Rewriting the variation (\ref{variation}) as
$$
\delta S=\int_{0}^{T}dt\int_{-\infty}^{\infty}dx\,\rho \,(\partial_{t}\delta h-\nu \partial_{x}^2\delta h -\lambda \partial_x h \,\partial_x \delta h),
$$
and integrating by parts, we arrive at the second Hamilton equation:
\begin{equation}\label{peqA}
\partial_{t}\rho=-\nu \partial_{x}^2 \rho +\lambda \partial_x \left(\rho \partial_x h\right).
\end{equation}
The boundary terms in $x$, emerging in the integrations by parts, vanish because of the boundary conditions at $|x|\to \infty$.  There also appear two boundary terms in time: at $t=0$ and $t=T$. The boundary term  $\int dx  \,\rho(x,0) \,\delta h(x,0)$ vanishes because the height profile at $t=0$ is fixed by Eq.~(\ref{parabola}). The boundary term $\int dx  \,\rho(x,T) \,\delta h(x,T)$ must be also zero. As we fixed $h(x=0,T)=H$, we have $\delta h(x=0,T) =0$, but $\rho(x=0,T)$ can be arbitrary. On the contrary,  $h(x\neq 0,T)$ is not fixed, so we must have $\rho(x\neq 0,T)=0$.  This leads to the boundary condition \cite{KK2007,MKV}
\begin{equation}\label{pTA}
    \rho(x,T)=\Lambda \,\delta(x),
\end{equation}
where one introduces an unknown constant $\Lambda$ which is ultimately set by the condition $h(x=0,T)=H$. Upon the rescaling $t/T\to t$, $x/\sqrt{\nu T} \to x$, $|\lambda|h/\nu\to h$ and  $|\lambda|T\rho/\nu\to \rho$, one arrives
at Eqs.~(\ref{eqh})-(\ref{action1}) of the main text, with rescaled $H$ and $\Lambda$, and Eq.~(\ref{parabola}) with rescaled $L$.


\begin{thebibliography}{99}
\bibitem{signlambda}
Changing the sign of $\lambda$ is  equivalent to changing the sign of $h$.
\bibitem{units} In the notation used in Ref. \cite{DMRS},
the large deviation function $\Phi(H)=\frac{1}{8} S(-2H)$, where $S$ is our action.
\end{thebibliography}
\end{document}